\documentclass[final,5p,times,twocolumn]{elsarticle}

\usepackage{subfig}
\usepackage{graphicx}
\usepackage{dcolumn}
\usepackage{bm}
\usepackage{amsfonts}
\usepackage{slashed}
\usepackage{mathtools}
\usepackage{amsmath}
\usepackage{xcolor}
\usepackage{soul}
\usepackage{float}
\usepackage[bottom]{footmisc}
\usepackage{amssymb}
\usepackage{lipsum}

\journal{Physics Letters B}

\begin{document}

\begin{frontmatter}


\title{Axial Anomaly and Confinement in Two-dimensional QED:\\Singular Behavior of Vacuum Polarization at Threshold}


\author[label1]{Bailing Ma}
\author[label2]{Chueng-Ryong Ji}
\affiliation[label1]{organization={School of Medicine, Wake Forest University},
            addressline={}, 
            city={Winston-Salem},
            postcode={27101}, 
            state={NC},
            country={USA}}
\affiliation[label2]{organization={Department of Physics and Astronomy, North Carolina State University},
            addressline={}, 
            city={Raleigh},
            postcode={27695-8202}, 
            state={NC},
            country={USA}}

\begin{abstract}
Performing the perturbative calculation of the vacuum polarization amplitude in $\text{QED}_{1+1}$, one finds an anomalous axial vector Ward identity. 
We note that the photon self-energy function displays a singularity in the 1+1D case, in stark contrast to the 3+1D case.
We discuss the nature of this singularity and 
its physical implications from the perspectives of axial anomaly and confinement
in two-dimensional QED. Computing the total cross section of $e^+e^-\to \mu^+\mu^-$
and displaying the toy version of the R ratio in 1+1D with respect to the typical R ratio in 3+1D, we discuss the significance of using the dressed photon propagator in obtaining the finite cross section.
\end{abstract}




\end{frontmatter}




\section{\label{sec:intro}Introduction}

The Schwinger model, or QED(1+1), provides a useful tool in understanding the confinement mechanism of fermions intuitively in the paradigm of relativistic quantum gauge field theories. This may be the reason why the confinement mechanism is frequently coined as    
the Schwinger mechanism. Because the gauge field lines are restricted to one dimension, the Schwinger model naturally provides a potential which confines fermions. When a fermion-antifermion pair is separated in 1+1D, it is energetically favorable to produce another fermion-antifermion pair out of the vacuum, shielding the originally separated fermion and antifermion not to yield any liberated fermion or anti-fermion. Such a mechanism of confinement (i.e. the ``Schwinger mechanism'') is tied to the intrinsic axial anomaly of the Dirac vacuum.  

While the axial anomaly in 3+1D is described by the triangle diagram amplitude proportional to the totally antisymmetric Levi-Civita symbol $\epsilon_{\mu\nu\alpha\beta}$, such an antisymmetric symbol cannot exist in 1+1D. However, the axial anomaly in 1+1D can still be described by the two-point amplitude proportional to the lowest rank antisymmetric tensor $\epsilon_{\mu\nu}$. This is consistent with the fact that the magnetic field is absent in 1+1D but the electric field can still be applied to the Dirac vacuum in 1+1D and excite a negative energy electron from the Dirac sea to a positive energy level. For this reason, the pair creation of electron and positron can be regarded as the phenomenon of axial anomaly in quantum field theory. The level crossing phenomenon in 3+1D occurs essentially the same way as the electric field applies to the same direction of the magnetic field which sets the level density of the Dirac vacuum in 3+1D. 

While the axial anomaly amplitude $T_5^{\mu\nu}$ is related to the one-loop vacuum polarization amplitude $T^{\mu\nu}$, namely  $T_5^{\mu\nu}=\epsilon^{\mu}_\lambda T^{\lambda\nu}$, the mass generation of the photon in 1+1D may be understood as modifying the free photon propagator by taking a geometric sum of the fermion loops in the propagation of the photon via the photon interaction with the Dirac vacuum. 
This mechanism of photon mass generation in 1+1D is distinct from the Higgs mechanism which provides the gauge boson mass by absorbing the Goldstone bosons in the 3+1D gauge field theory, as is well known in the electroweak unification of the Standard Model. In 1+1D, the spontaneous symmetry breaking cannot accompany the Goldstone boson according to Coleman's theorem~\cite{Coleman:1973ci}. 

However, in 1+1D gauge field theory, one may study the confinement mechanism as well as the mechanism of photon mass generation in the purview of the axial anomaly phenomenon intrinsic to the Dirac vacuum. There have been various discussions of the axial anomaly regarding the contradistinction between the two dimensional QED and the four dimensional QCD~\cite{Pak:1977an, Callan:1976je, tHooft:1976rip,Jackiw:1976pf}. It has been also noted that the axial anomaly in 1+1D may correspond to the Chern-Simons term in AdS$_3$ holographic theory~\cite{Katz:2007br}.
In Ref.~\cite{Katz:2007br}, the authors discussed the dual correspondence between the 't Hooft model (i.e. QCD$_{1+1}$ in the large $N_c$ limit) and the AdS$_3$ holographic model, as well as the features of axial anomaly from the perspectives of conformal symmetry. 
Due to the absence of the axial anomaly in non-abelian $SU(N_c)$ theories, one may regard this correspondence from the original 't Hooft gauge group $U(N_c)$ in the limit of $N_c \to \infty$ with the correspondence of 
$U(N_c) \approx SU(N_c-1) \otimes U(1)$. 
In this respect, essentially the same AdS$_3$ holographic correspondence can be applied to the Schwinger model, or QED(1+1), as well. In this work, we do not provide any further remarks on the holographic correspondence or bosonization analyses of the Schwinger model, but we  
discuss the axial anomaly of the Schwinger model from the perspective of the genuine 1+1D gauge field theory.

Before we lay out the details of our work, we first briefly summarize the more familiar formal derivation of photon mass generation from the perspective of axial anomaly. To illustrate axial anomaly more heuristically in terms of the Dirac sea, one may consider the massless Schwinger model on a circle, with the fields satisfying the boundary conditions
\begin{equation}
    A_{\mu}\left(t,x=-\frac{L}{2}\right)=A_{\mu}\left(t,x=\frac{L}{2}\right),
\end{equation}
\begin{equation}
    \psi\left(t,x=-\frac{L}{2}\right)=-\psi\left(t,x=\frac{L}{2}\right).
\end{equation}
Since the Lagrangian is invariant under $U_V(1)$ and $U_A(1)$ symmetries, classically both the charge $Q(t)$ and axial-charge $Q^5(t)$ defined by   
\begin{equation}
    Q(t)=\int dx J_0(t,x)
\end{equation}
and
\begin{equation}
    Q^5(t)=\int dx J_0^5(t,x)
\end{equation}
are conserved. As $Q(t)$ and $Q^5(t)$ may be decomposed in terms of the left-moving and right-moving fermions, i.e. 
\begin{equation}
    Q(t) = Q_L(t) + Q_R(t)
\end{equation}
and
\begin{equation}
    Q^5(t) = Q_L(t) - Q_R(t),
\end{equation}
the left-moving charge $Q_L(t)$ and the right-moving charge $Q_R(t)$ are separately conserved as well. However, this is no longer true in the quantum field theory due to the nature of the Dirac sea. In the Coulomb gauge, the electric field is given by the time derivative of the gauge field $A_1(t,x)$.
Let us say the Dirac sea is provided by filling up all the negative energy levels and all the positive energy levels are empty at the zero gauge field, i.e. $A_1=0$. Now, applying an external electric field $\partial A_1(t,x)/\partial t$, one may increase $A_1$ from $0$ to $2\pi/L$, creating a pair of L-particle and R-hole due to the response of both left-moving and right-moving fermions in the Dirac sea with respect to the applied external electric field. Such particle-hole pair creation under the external electric field may illustrate the axial anomaly in 1+1D. The charge is still conserved since the electric charge of the particle and the hole are opposite. The axial charges, however, are identical for both so that the total axial charge changes by 2 units, namely
$\Delta Q^5 = \frac{eL}{\pi} \Delta A_1 = \frac{eL}{\pi} \frac{2\pi}{L}=2e$.
We may summarize this famous anomaly result as
\begin{equation}
\label{axial-anomaly}
    \partial^{\mu}J_{\mu}^5=\frac{e}{\pi}\varepsilon_{\mu\nu}\partial^{\mu}A^{\nu},
\end{equation}
in contrast to the continuity equation of the vector current
\begin{equation}
    \partial^{\mu}J_{\mu}=0.
\end{equation}
In 1+1D, one may surmise the relationship between the axial-vector current and the vector current given by
\begin{equation}
\label{Axial-Vector}
    J^{5}_{\mu}=\varepsilon_{\mu\nu}J^{\nu}.
\end{equation}
Realizing the off-diagonal element of the 1+1D field strength tensor $F^{\mu\nu}$
as the electric field $E(t,x)$, we may write 
\begin{equation}
\label{E-field}
    F^{\mu\nu}=\varepsilon^{\mu\nu} E,
\end{equation}
where $F^{\mu\nu}$ satisfies the Maxwell's equation or the Gauss's law given by
\begin{equation}
\label{Gauss-law}
\partial_{\mu}F^{\mu\nu}=e\bar{\psi}\gamma^{\nu}\psi=eJ^{\nu}.
\end{equation}
From Eq. (\ref{axial-anomaly}) as well as Eqs. (\ref{Axial-Vector})-(\ref{Gauss-law}), one can find the following equation 
\begin{equation}
   e \, \partial^{\mu}J^5_{\mu}=-\partial^\mu \partial_\mu E= \frac{e^2}{2\pi}\varepsilon_{\mu\nu}F^{\mu\nu} = -\frac{e^2}{\pi}E ,
\end{equation}
where $\epsilon_{\mu\nu}  \epsilon^{\alpha\nu}  = - g_\mu^\alpha$ is used.
This derivation from the axial anomaly given by Eq.~(\ref{axial-anomaly}) provides the Klein-Gordon equation for the electric field $E(t,x)$ given by 
\begin{equation}
\label{axial-anomaly-result}
    \Box E =\frac{e^2}{\pi}E.
\end{equation}
Here, the mass of photon can be identified as $\frac{e}{\sqrt{\pi}}$, illustrating photon mass generation from the axial anomaly. The same result in Euclidean space was discussed in Ref~\cite{Adam:1993fy} and
the corresponding second order differential equation for the axial charge $Q^5(t)$ was discussed~\cite{Pak:1977an} in conjunction with the understanding of the $U_{A}(1)$ problem in the $\theta$-vacuum.   
 
In this paper, we will discuss the symptoms of confinement in QED(1+1) looking into the result of the fermion one-loop amplitude associated with the axial anomaly and analyzing its dispersion relation. We will 
compare this result with the result in QED(3+1) and discuss the difference from the perspectives of confinement in 1+1D vs. deconfinement in 3+1D. 

This paper is organized as follows. In Sec.~\ref{sec:selfenergy}, we present the one loop self-energy of the photon in 1+1D and 3+1D and show that their dispersion relations are both satisfied. We show that a singularity in the photon self-energy function is present in the 1+1D case but not 3+1D. We relate this feature to the confinement mechanism. In Sec.~\ref{sec:allloop}, we present the exact photon propagator as a geometric sum of the one-particle-irreducible insertions(1PI), where the 1PI considered is up to one-loop order. The singularity discussed previously disappears when summing over all 1PI. In Sec.~\ref{sec:physical} we compute physical observables such as the total cross section of $e^+e^-\rightarrow \mu^+\mu^-$, and the R-ratio, comparing the 1+1D and 3+1D cases. We summarize our conclusions in Sec.~\ref{sec:conclusion}. In~\ref{sec:app}, we contrast the ultra-relativistic limits of the total cross sections in 1+1D and 3+1D, noting the contributions from longitudinal and transverse photon polarizations, respectively.

\section{\label{sec:selfenergy} Vacuum polarization and its dispersion relation }
\begin{figure}
	\centering
\includegraphics[width=0.75\linewidth]{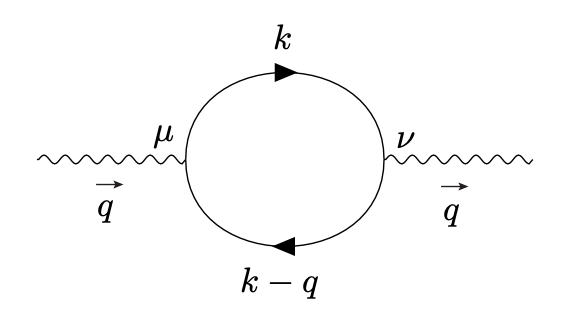}
	\caption{Feynman diagram for the photon self-energy at one-loop order.}
	\label{fig:PhotonSE}
\end{figure}
As the axial anomaly in two-dimensional quantum electrodynamics ($\text{QED}_2$) is proportional to the lowest rank anti-symmetric tensor $\epsilon_{\mu\nu}$, it cannot be represented by the typical anomalous triangle diagram proportional to the Levi-Civita tensor $\epsilon_{\mu\nu\alpha\beta}$ as in 3+1D quantum field theories such as $\text{QED}_4$ and $\text{QCD}_4$, but can be described by the two-point function \cite{Chen1999}, as shown in Fig.~\ref{fig:PhotonSE}. 
The covariant calculation of the Feynman diagram shown in Fig.~\ref{fig:PhotonSE} was discussed in~\cite{Adam:1993fy}. The amplitude with vector and axial vector current ($\gamma^{\mu}$ and $\gamma^{\nu}\gamma_5$ instead of $\gamma^{\mu}$ and $\gamma^{\nu}$ in Fig.~\ref{fig:PhotonSE}) can be related to the vacuum polarization amplitude by
\begin{equation}\label{eqn:axialrelatedtovector}
T^{\mu\nu}_5=\varepsilon^{\nu\lambda}T_{\lambda}^{\mu},
\end{equation}
due to the relationship $\gamma^{\mu}\gamma^5=\varepsilon^{\mu\nu}\gamma_{\nu}$ in $1+1$ dimensions.

For the discussion of the dispersion relation, we may briefly summarize the vacuum polarization diagram shown in Fig.~\ref{fig:PhotonSE} \footnote{Detailed time-ordered computation corresponding to this covariant axial anomaly amplitude will be presented in our forthcoming paper.}, i.e. 
\begin{align}
T^{\mu\nu}(q)=ie^2\int\frac{d^2k}{(2\pi)^2}\frac{Tr\left[\gamma^{\mu}(\slashed{k}+m)\gamma^{\nu}(\slashed{k}-\slashed{q}+m)\right]}{\left[k^2-m^2+i\epsilon\right]\left[(k-q)^2-m^2+i\epsilon\right]},\label{eqn:Tmunuaxial}
\end{align}
which is given by 
\begin{align}\label{eqn:anomalyTmunu}
&T^{\mu\nu}(q)\notag\\
=&-\frac{e^2}{2\pi}\int_0^1 dx \frac{x(x-1)(2q^{\mu}q^{\nu}-g^{\mu\nu}q^2)+g^{\mu\nu}m^2}{x(x-1)q^2+m^2}+\frac{e^2}{2\pi}g^{\mu\nu},
\end{align}
using the Feynman parametrization.
It manifestly satisfies the gauge invariance
\begin{equation}
    q_{\mu}T^{\mu\nu}(q)=q_{\nu}T^{\mu\nu}(q)=0.
\end{equation}
Indeed, $T^{\mu\nu}(q)$ has the tensor structure of 
\begin{equation}
T^{\mu\nu}(q)=T(q^2)\left(q^{\mu}q^{\nu}-g^{\mu\nu}q^2\right).\label{eqn:anomalyTmunugaugeinv}
\end{equation}
Using Eq. (\ref{eqn:axialrelatedtovector}), one may write
\begin{align}
T^{\mu\nu}_5(q)&=T(q^2)\varepsilon^{\nu\lambda}\left(q^{\mu}q_{\lambda}-g^{\mu}_{\lambda}q^2\right),
\end{align}
fulfilling the vector current conservation,
\begin{align}
q_{\mu}T^{\mu\nu}_5(q)&=T(q^2)\varepsilon^{\nu\lambda}q_{\mu}\left(q^{\mu}q_{\lambda}-g^{\mu}_{\lambda}q^2\right)=0,
\end{align}
as well as the anomalous axial vector current,
\begin{align}\label{eqn:anomalousAWI}
q_{\nu}T^{\mu\nu}_5(q)&=T(q^2)\varepsilon^{\nu\lambda}q_{\nu}\left(q^{\mu}q_{\lambda}-g^{\mu}_{\lambda}q^2\right)=-q_{\nu}\varepsilon^{\nu\mu}q^2T(q^2).
\end{align}

If one computes the two-point function with a vector and a pseudo-scalar current,
\begin{equation}
P^{\mu}_5(q)=ie^2\int\frac{d^2k}{(2\pi)^2}\frac{Tr\left[\gamma^{\mu}(\slashed{k}+m)\gamma_{5}(\slashed{k}-\slashed{q}+m)\right]}{\left[k^2-m^2+i\epsilon\right]\left[(k-q)^2-m^2+i\epsilon\right]},
\end{equation}
one gets
\begin{equation}\label{eqn:twopointvectorpseudoscalar}
P^{\mu}_5(q)=\frac{e^2m}{2\pi}\int_0^1 dx \frac{\varepsilon^{\mu\nu}q_{\nu}}{x(x-1)q^2+m^2}.
\end{equation}
Thus, one can write
\begin{equation}
q_{\nu}T^{\mu\nu}_5(q)=2m P^{\mu}_5(q)+\frac{e^2}{\pi}q_{\lambda}\varepsilon^{\lambda\mu}.
\end{equation}
This is the famous axial anomaly which 
reduces to
\begin{align}\label{eqn:anomalousAWImassless}
q_{\nu}T^{\mu\nu}_5(q)&=\frac{e^2}{\pi}q_{\nu}\varepsilon^{\nu\mu}
\end{align}
in the massless case. 
As has long been known, one cannot conserve both axial vector and vector currents~
\cite{Sutherland:1966zz, Sutherland:1967vf}.

After the integration of the Feynman parameter $x$, the explicit form of the $T(q^2)$ function in $1+1$ dimensions is given by~\cite{Adam:1993fy}
\begin{align}\label{eqn:resTq2int}
T(q^2)
=-\frac{e^2}{\pi q^2}\left[1-\frac{m^2/q^2}{\sqrt{1/4-m^2/q^2}}\ln\left(\frac{1-\frac{1}{2\sqrt{1/4-m^2/q^2}}}{1+\frac{1}{2\sqrt{1/4-m^2/q^2}}}\right)\right].
\end{align}
While we confirm this result~\cite{Adam:1993fy},we note that this result indicates an interesting difference above and below the threshold value $q^2=4m^2$. The process of a pair creation of fermion and anti-fermion can occur above the threshold, $q^2>4m^2$, while no such process can be allowed below the threshold, $q^2<4m^2$. Indeed, the characteristic of the amplitude $T(q^2)$ is quite different between the two regions of $q^2$ above and below the threshold. 
On one hand, for $q^2>4 m^2$, the real and imaginary parts of the amplitude $T(q^2)$ are given by 
\begin{align}\label{eqn:resTq2large}
&\text{Re}\left[T(q^2)\right]\notag\\
=&-\frac{e^2}{\pi q^2}\left[1-\frac{m^2/q^2}{\sqrt{1/4-m^2/q^2}}\ln\left(\frac{\frac{1}{2\sqrt{1/4-m^2/q^2}}-1}{\frac{1}{2\sqrt{1/4-m^2/q^2}}+1}\right)\right]
\end{align}
and
\begin{align}
\text{Im}\left[T(q^2)\right]=\frac{e^2}{q^2}\frac{m^2/q^2}{\sqrt{1/4-m^2/q^2}}\Theta(q^2-4m^2).
\end{align}
On the other hand, for $q^2<4 m^2$, the amplitude $T(q^2)$ is purely real as given by
\begin{align}\label{eqn:resTq2small}
T(q^2)=-\frac{e^2}{\pi q^2}\left[1-2\frac{m^2/q^2}{\sqrt{m^2/q^2-1/4}}\tan^{-1}\left(\frac{1}{2\sqrt{m^2/q^2-1/4}}\right)\right].
\end{align}
Thus, we notice here a remarkable singular behavior of the vacuum polarization function $T(q^2)$ at $q^2=4m^2$. We interpret this singular behavior at $q^2=4m^2$ as the symptom of the confinement in QED(1+1), namely a fermion or anti-fermion cannot be on its mass shell, \'a la ``Schwinger mechanism''.

We may contrast this singular behavior of the vacuum polarization amplitude in 1+1D with the 3+1D vacuum polarization amplitude $T_{3+1}(q^2)$ given by~\cite{BjorkenDrell} 
\begin{align}
    &T_{3+1}(q^2)\notag\\
    =&-\frac{\alpha}{3\pi}\left[\log\frac{\Lambda_{PV}^2}{m^2}-6\int_0^1 dx\, x(1-x)\log\left(1-\frac{q^2}{m^2}x(1-x)\right)\right],
\end{align}
where $\Lambda_{PV}$ is the Paul-Villar's regularization parameter for the regularization of UV-divergence in 3+1D. The regularization scale independent formula may also be defined as 
\begin{align}\label{eqn:That3+1}
    &\hat{T}_{3+1}(q^2)\equiv T_{3+1}(q^2)-T_{3+1}(0)\notag\\
    =&\frac{2\alpha}{\pi}\int_0^1 dx\, x(1-x)\log\left(\frac{m^2}{m^2-x(1-x)q^2}\right).
\end{align}
Apparently, the vacuum polarization amplitudes 
$T(q^2)$ in 1+1D and $T_{3+1}(q^2)$ or $\hat{T}_{3+1}(q^2)$ behave far differently near the on-mass-shell threshold value of $q^2=4m^2$ for the pair creation of fermion and anti-fermion. 

We verify that both $T(q^2)$ in 1+1D and $\hat{T}_{3+1}(q^2)$ satisfy the dispersion relation (DR), respectively. In 1+1D, $T(q^2)$ satisfies the DR given by 
\begin{align}\label{eqn:DR}
    \text{Re}T(q^2)&=\frac{1}{\pi}P\int_{-\infty}^{+\infty}\frac{\text{Im}T(q'^2)}{q'^2-q^2}dq'^2,\notag\\
    \text{Im}T(q^2)&=-\frac{1}{\pi}P\int_{-\infty}^{+\infty}\frac{\text{Re}T(q'^2)}{q'^2-q^2}dq'^2,
\end{align}
where $P$ indicates the Cauchy principal value. In 3+1D, $\hat{T}_{3+1}(q^2)$ satisfies the subtracted DR given by
\begin{align}
    \text{Re}\hat{T}_{3+1}(q^2)&=\frac{1}{\pi}\left(P\int_{-\infty}^{+\infty}\frac{\text{Im}\hat{T}_{3+1}(q'^2)}{q'^2-q^2}dq'^2\right.\notag\\
    &\left.-P\int_{-\infty}^{+\infty}\frac{\text{Im}\hat{T}_{3+1}(q'^2)}{q'^2-q_0^2}dq'^2\right)\notag\\
    &=\frac{q^2-q_0^2}{\pi}P\int_{-\infty}^{+\infty}\frac{\text{Im}\hat{T}_{3+1}(q'^2)}{(q'^2-q^2)(q'^2-q_0^2)}dq'^2,\notag\\
    \text{Im}\hat{T}_{3+1}(q^2)&=-\frac{1}{\pi}\left(P\int_{-\infty}^{+\infty}\frac{\text{Re}\hat{T}_{3+1}(q'^2)}{q'^2-q^2}dq'^2\right.\notag\\
    &\left.-P\int_{-\infty}^{+\infty}\frac{\text{Re}\hat{T}_{3+1}(q'^2)}{q'^2-q_0^2}dq'^2\right)\notag\\
    &=-\frac{q^2-q_0^2}{\pi}P\int_{-\infty}^{+\infty}\frac{\text{Re}\hat{T}_{3+1}(q'^2)}{(q'^2-q^2)(q'^2-q_0^2)}dq'^2,
\end{align}
where $q_0^2$ is taken to be $0$ to cross the origin at $q^2=0$ in agreement with Eq.~(\ref{eqn:That3+1}).

\begin{figure}
	\centering
    \subfloat[]{\includegraphics[width=0.95\linewidth]{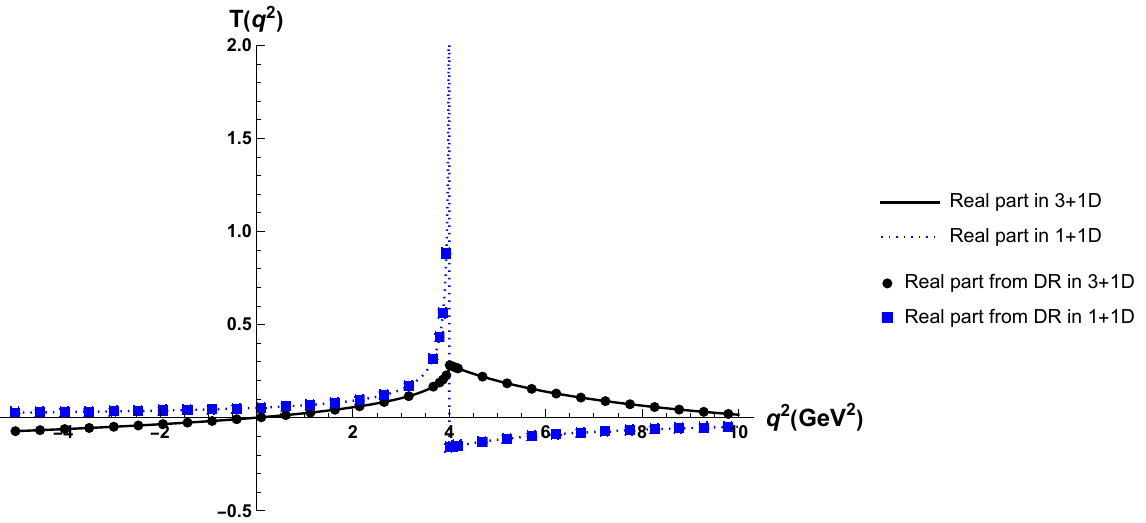}\label{fig:PlotTqsq_DR_Re}}\\
    \subfloat[]{\includegraphics[width=0.95\linewidth]{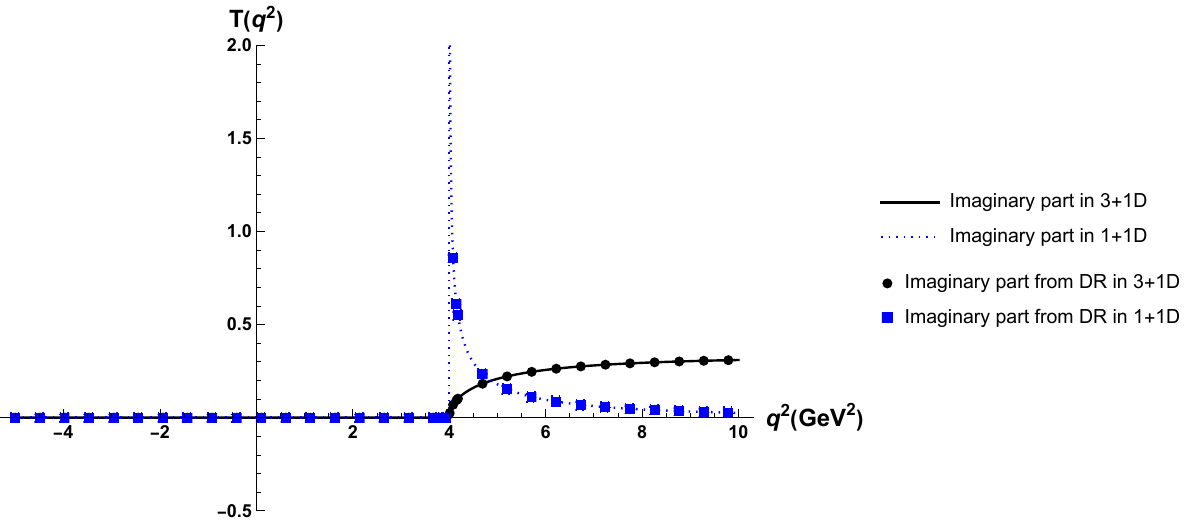}\label{fig:PlotTqsq_DR_Im}}
	\caption{Plot of (a) $\text{Re}[T(q^2)]$ and (b) $\text{Im}[T(q^2)]$ in 1+1D case taking $m=1$ and $e=1$, and in 3+1D case taking $m=1$ and $\alpha=1$, where the dots represent results from the (subtracted) dispersion relation.}
	\label{fig:PlotTqsq_DR_Re_Im}
\end{figure}

 In Fig.~\ref{fig:PlotTqsq_DR_Re_Im}, the real and imaginary parts of $T(q^2)$ in both 1+1D and 3+1D are plotted, where the data points represent the DR result while the lines represent the direct result. For the plots in the figures, we take $m$ and $e$ with proper units, e.g. $m$ in GeV for both 1+1D and 3+1D while $e$ in GeV for 1+1D case only.
 We can see in both cases the (unsubtracted or subtracted) DR is satisfied. 
 In the 1+1D case, however, both the real and imaginary parts diverge at $q^2=4m^2$, 
indicating no physical pair production of on-mass-shell particle and anti-particle at the $q^2=4m^2$ threshold, while in 3+1D case there is no such abnormal behavior of the amplitudes indicating physical pair production of on-mass-shell particle and anti-particle at the threshold and beyond. In 3+1D case, both functions of $\text{Re}[T(q^2)]$ and $\text{Im}[T(q^2)]$ are not only connected at $q^2=4m^2$ but indeed keep their values finite for the entire time-like region above the threshold, $q^2>4m^2$. As $q^2$ gets larger, $\text{Re}[T(q^2)]$ grows in its value while $\text{Im}[T(q^2)]$ saturates to a finite value. In contrast, in 1+1D case, there is not only a jump in value, for $\text{Re}[T(q^2)]$ from $+\infty$ to $-\frac{1}{2\pi}$ from the left of $q^2=4m^2$ to the right of it, while for $\text{Im}[T(q^2)]$ from $0$ to $+\infty$, but also both $\text{Re}[T(q^2)]$ and $\text{Im}[T(q^2)]$ die out to zero as $q^2$ gets larger. This indicates that no real on-mass-shell particle and anti-particle pair productions occur in 1+1D although virtual off-mass-shell particle and anti-particle pair productions may occur within the limits set by the quantum mechanical uncertainty principle. 

\section{\label{sec:allloop} Photon mass generation in 1+1D}

\begin{figure}
	\centering
\includegraphics[width=1.0\linewidth]{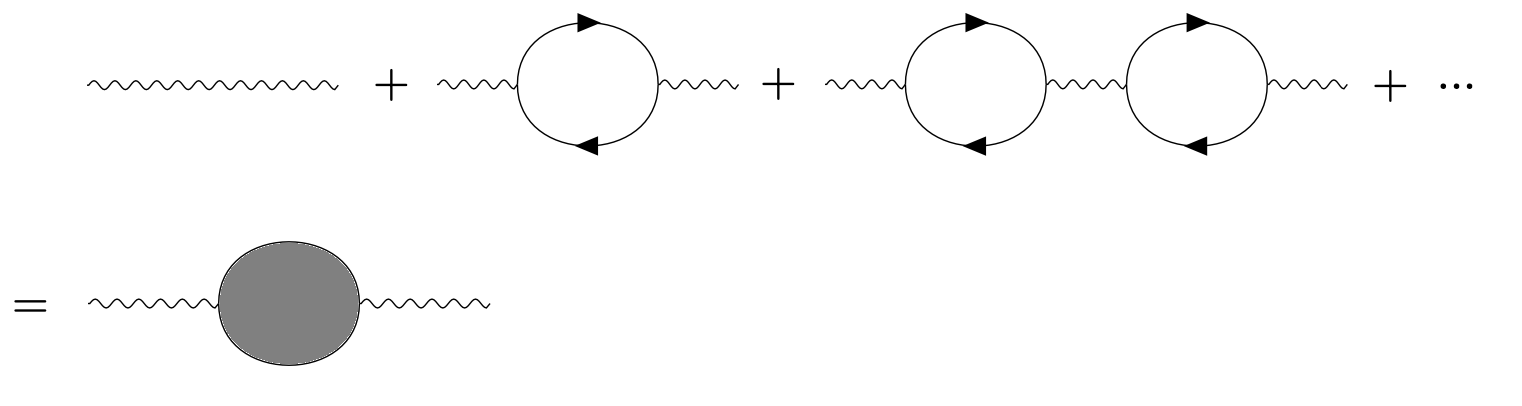}
	\caption{Mass gap equation for the photon in $\text{QED}_{1+1}$.}
	\label{fig:Photon_dress}
\end{figure}

The photon mass generation in 1+1D may be understood from the interaction of the propagating photon with the Dirac vacuum which leads to the photon self-energy amplitude as shown in Fig.~\ref{fig:PhotonSE}
at the lowest order. In the previous section, Sec.~\ref{sec:selfenergy}, we discussed the axial anomaly in QED$_2$ using the same amplitude (Fig.~\ref{fig:PhotonSE}).
As discussed in Ref.~\cite{Adam:1993fy}, the geometric sum of photon self-energy process may lead to the photon mass gap amplitude shown in Fig.~\ref{fig:Photon_dress}. 
We note that the geometrical sum of the one-loop $T(q^2)$, i.e. $\frac{1}{1+T(q^2)}$, removes the singularity at $q^2=4 m^2$. The photon mass can also be immediately seen in the result of the geometric sum as illustrated below. 

While the bare photon propagator may be denoted as $\left(g_{\mu\nu}-\frac{q_{\mu}q_{\nu}}{q^2}\right)D_0(q^2)$ in the Lorentz gauge, where
\begin{equation}
    D_0(q^2)=-\frac{1}{q^2+i\epsilon},\label{eqn:photon_freeprop}
\end{equation}
the dressed photon propagator may be summarized as
\begin{equation}
    D_{\mu\nu}(q)=\left(g_{\mu\nu}-\frac{q_{\mu}q_{\nu}}{q^2}\right)D(q^2)\label{eqn:photon_prop}.
\end{equation}
Denoting the dressed photon propagator (Fig.~\ref{fig:Photon_dress}), we get
\begin{figure}
	\centering
    \subfloat[]{\includegraphics[width=0.95\linewidth]{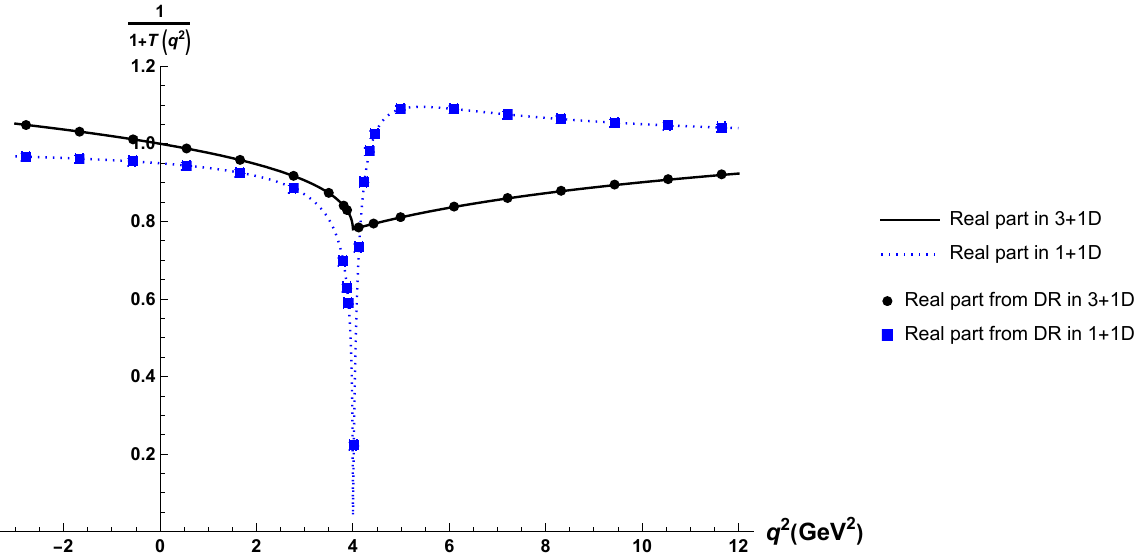}\label{fig:PlotdreTqsq_DR_Re}}\\
    \subfloat[]{\includegraphics[width=0.95\linewidth]{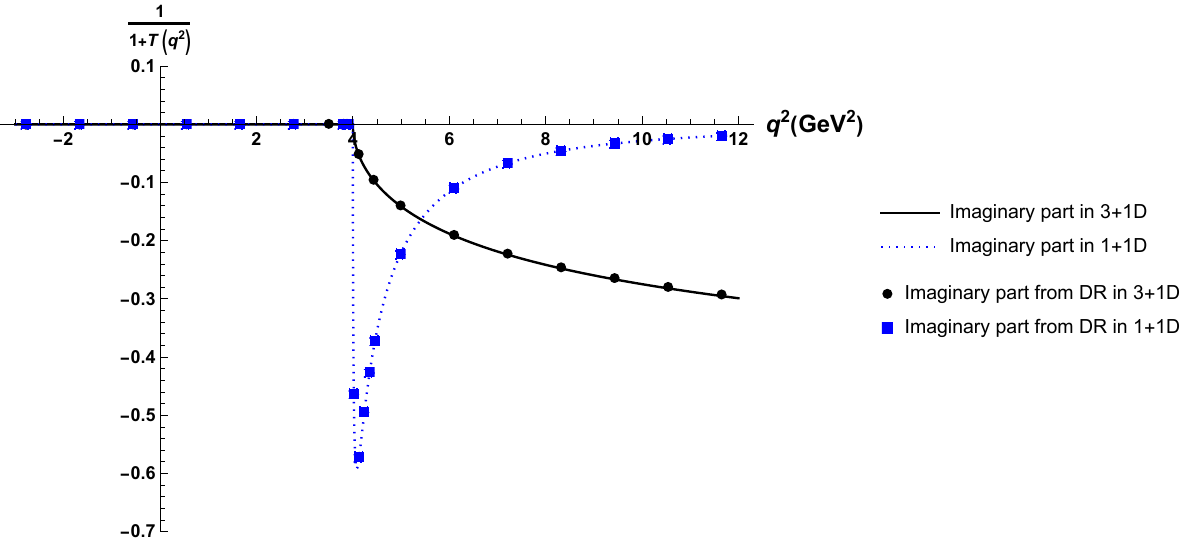}\label{fig:PlotdreTqsq_DR_Im}}
	\caption{Plot of (a) $\text{Re}[\frac{1}{1+T(q^2)}]$ and (b) $\text{Im}[\frac{1}{1+T(q^2)}]$ in 1+1D and 3+1D, where the dots represent results from the subtracted dispersion relation.}
	\label{fig:PlotdreTqsq_DR_Re_Im}
\end{figure}
    \begin{align}
        &i\left(g_{\mu\nu}-\frac{q_{\mu}q_{\nu}}{q^2}\right)D_0(q^2)\notag\\
        &+i\left(g_{\mu\nu}-\frac{q_{\mu}q_{\nu}}{q^2}\right)D_0(q^2)\ iT^{\mu\nu}(q^2)\ i\left(g_{\mu\nu}-\frac{q_{\mu}q_{\nu}}{q^2}\right)D_0(q^2)\notag\\
        &+i\left(g_{\mu\nu}-\frac{q_{\mu}q_{\nu}}{q^2}\right)D_0(q^2)\ iT^{\mu\nu}(q^2)\ i\left(g_{\mu\nu}-\frac{q_{\mu}q_{\nu}}{q^2}\right)D_0(q^2)\notag\\
        &\cdot iT^{\mu\nu}(q^2)\ i\left(g_{\mu\nu}-\frac{q_{\mu}q_{\nu}}{q^2}\right)D_0(q^2)+...\notag\\
        =&\ i\left(g_{\mu\nu}-\frac{q_{\mu}q_{\nu}}{q^2}\right)D(q^2),
    \end{align}
where we can readily obtain
\begin{align}
    D(q^2)
    =-\frac{1}{q^2+q^2T(q^2)}.\label{eqn:photon_dressedprop}
\end{align}
In the case of massless fermions ($m=0$), we use Eq.~(\ref{eqn:resTq2int}) to obtain
\begin{equation}
    D_{\mu\nu}(q)=-\frac{g_{\mu\nu}-q_{\mu}q_{\nu}/q^2}{q^2-e^2/\pi+i\epsilon}.
\end{equation}
This provides the photon propagator with mass $m_{\gamma}=e/\sqrt{\pi}$.
Moreover, the geometric sum shown in Fig.~\ref{fig:Photon_dress} eliminates the singularity of the vacuum polarization amplitude $T(q^2)$ at $q^2=4m^2$ present in the individual diagrams in the first line of Fig.~\ref{fig:Photon_dress}. As
\begin{equation}
    D(q^2)=\frac{1}{1+T(q^2)}D_0(q^2),
\end{equation}
and the plot of $\frac{1}{1+T(q^2)}$ is shown in Fig.~\ref{fig:PlotdreTqsq_DR_Re_Im}, we can see that both the real and imaginary parts of $\frac{1}{1+T(q^2)}$ are finite everywhere, and also they satisfy the dispersion relation (DR).

In comparison to the cut diagram of Fig.~\ref{fig:PhotonSE}, the cut diagram for the blobbed graph in Fig.~\ref{fig:Photon_dress} represents effectively the pair production of 
hadronized final particles instead of bare particles.  
As the hadronized final particles are not confined, the corresponding amplitude is finite as shown in Fig.~\ref{fig:PlotdreTqsq_DR_Re_Im}.

\section{Physical observables: cross section and R-ratio\label{sec:physical}}
\begin{figure}
	\centering
\includegraphics[width=0.95\linewidth]{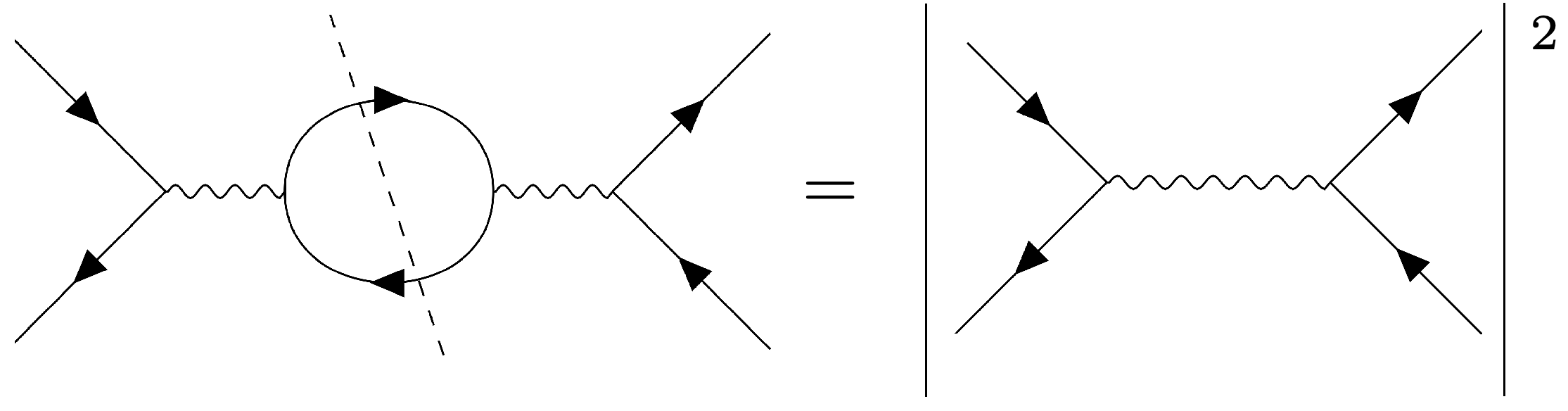}
\caption{Total cross-section of $e^+e^-\rightarrow \mu^+\mu^-$.}
	\label{fig:cross-section}
\end{figure}

In $3+1$ dimensions, the scattering cross section of the $e^+ e^- \rightarrow \mu^+ \mu^-$ process is an exemplary physical observable. In the leading order of the QED coupling constant, the total unpolarized  cross section of $e^+e^-\rightarrow \mu^+\mu^-$ (See Fig.~\ref{fig:cross-section}) is given by 
\begin{align}
\label{cross-section(3+1)}
    &\sigma_{\text{tot},3+1}^{e^+e^-\rightarrow \mu^+\mu^-}\notag\\
    =&\frac{1}{2w_e}\frac{1}{(2\pi)^2}\int \frac{d^3p_1^{'}}{2p_1^{'0}}\int \frac{d^3p_2^{'}}{2p_2^{'0}}\delta^{(4)}(p_1+p_2-p_1'-p_2')\ |\overline{\mathcal{M}}_{3+1}|^2,
\end{align}
where 
$
    w_e \equiv q^2\sqrt{1-\frac{4m_e^2}{q^2}}
$
and
\begin{align}
\label{Invariant-Amp(3+1)}
|\overline{\mathcal{M}}_{3+1}|^2=\frac{8e^4}{q^4} &\left(p_1\cdot p_1'\ p_2\cdot p_2'+p_1\cdot p_2'\ p_2\cdot p_1'+p_1\cdot p_2 m_{\mu}^2\right.\notag\\
    &\left.+p_1'\cdot p_2' m_e^2+2m_e^2m_{\mu}^2\right),
\end{align}
using the photon propagator $D_0(q^2)=-\frac{1}{q^2+i\varepsilon}$.
Using the phase space integration $\int \frac{d^3p_2^{'}}{2p_2^{'0}}=\int d^4p_2^{'}\delta\left(p_2^{'2}-m_{\mu}^2\right)$, and taking the center-of-mass frame, i.e., $p_1^{'0}=p_2^{'0}=\frac{\sqrt{q^2}}{2}=\frac{\sqrt{s}}{2}$, one can find 
\begin{align}
    &\sigma_{\text{tot},3+1}^{e^+e^-\rightarrow \mu^+\mu^-}\notag\\
    =&\frac{1}{2w_e}\frac{1}{(2\pi)^2}\int dp_1^{'0}\frac{\sqrt{(p_1^{'0})^2-m_{\mu}^2}}{2}\frac{1}{2q^0}\delta(p_1^{'0}-\frac{q^2}{2q^0})d\Omega\ |\overline{\mathcal{M}}_{3+1}|^2.
\end{align}
One can obtain
\begin{align}\label{cross-section(3+1)res}
 &\sigma_{\text{tot},3+1}^{e^+e^-\rightarrow \mu^+\mu^-}=\frac{1}{64\pi^2s}\frac{\sqrt{1-\frac{4m_{\mu}^2}{q^2}}}{\sqrt{1-\frac{4m_e^2}{q^2}}}\int d\Omega\ |\overline{\mathcal{M}}_{3+1}|^2\Theta(q^2-4m_{\mu}^2),
\end{align}
where the Heaviside theta function is to ensure that $q^2>4m_{\mu}^2$. Because $m_{\mu}>m_e$, this automatically ensures $q^2>4m_e^2$ also.

In contrast, the result in 1+1D corresponding to Eqs.~(\ref{cross-section(3+1)})-(\ref{Invariant-Amp(3+1)}) is given by
\begin{align}
\label{sigma}
    &\sigma_{\text{tot},1+1}^{e^+e^-\rightarrow \mu^+\mu^-}\notag\\
    =&\frac{1}{2w_e}\int \frac{dp_1^{'1}}{2p_1^{'0}}\int \frac{dp_2^{'1}}{2p_2^{'0}}\delta^{(2)}(p_1+p_2-p_1'-p_2')\ |\overline{\mathcal{M}}_{1+1}|^2,
\end{align}
where
\begin{align}
&|\overline{\mathcal{M}}_{1+1}|^2\notag\\
=&
    \frac{2e^4}{q^4}(p_1\cdot p_1'\ p_2\cdot p_2'+p_1\cdot p_2'\ p_2\cdot p_1'-p_1\cdot p_2\ p_1'\cdot p_2'+m_e^2m_{\mu}^2).
\end{align}
The integration of the momentum variables $p'_1$ and $p'_2$ of Eq.(\ref{sigma}) can be done straightforwardly
using the formulae Eqs. (260) and (261) in Ref.~\cite{Adam:1993fy}, respectively,
\begin{align}\label{eqn:ref260mu}
    I_2&=\int \frac{dp_1^{'1}}{2p_1^{'0}}\int \frac{dp_2^{'1}}{2p_2^{'0}}\delta^{(2)}(q-p_1'-p_2')f(p_1'\cdot p_2')\notag\\
    &=\frac{1}{w_{\mu}}f\left(\frac{1}{2}(q^2-2m_{\mu}^2)\right)
\end{align}
and
\begin{align}\label{eqn:ref261mu}
    I_2^{\mu\nu}&=\int \frac{dp_1^{'1}}{2p_1^{'0}}\int \frac{dp_2^{'1}}{2p_2^{'0}}\delta^{(2)}(q-p_1'-p_2')f(p_1'\cdot p_2')p_1^{'\mu}p_2^{'\nu}\notag\\
    &=\left[\left(\frac{q^2}{4}-m_{\mu}^2\right)g^{\mu\nu}+\frac{m_{\mu}^2}{q^2}q^{\mu}q^{\nu}\right]I_2,
\end{align}
where 
\begin{align}
   w_{\mu} \equiv q^2\sqrt{1-\frac{4m_{\mu}^2}{q^2}}.
\end{align}
Taking $f(p_1'\cdot p_2')=\frac{1}{2w_e}\frac{2e^4}{q^4}\left(-p_1\cdot p_2\ p_1'\cdot p_2'+m_e^2m_{\mu}^2\right)$ for Eq. (\ref{eqn:ref260mu}) and $f(p_1'\cdot p_2')=1$ along with the external tensor $\frac{1}{2w_e}\frac{2e^4}{q^4}\left(p_{1\mu}p_{2\nu}+p_{2\mu}p_{1\nu}\right)$ outside the integration of the $p'_1$ and $p'_2$ variables for Eq. (\ref{eqn:ref261mu}), and noting that 
\begin{figure}
	\centering
\includegraphics[width=0.95\linewidth]{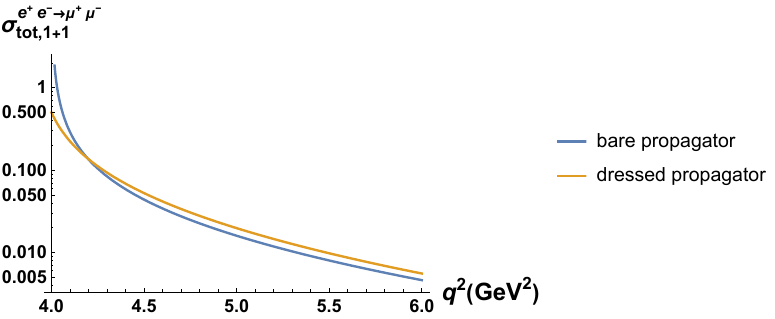}
	\caption{Total cross-section of $e^+e^-\rightarrow \mu^+\mu^-$ in 1+1D plotted as a function of $q^2$, using the bare propagator and the dressed propagator of the photon, respectively.}
	\label{fig:sigmatotaleeee}
\end{figure}
\begin{align}
    p_1'\cdot p_2'=\frac{q^2-2m_{\mu}^2}{2}\label{eqn:p1pdotp2p},
\end{align}
\begin{align}
    p_1\cdot p_2=\frac{q^2-2m_e^2}{2}\label{eqn:p1dotp2},
\end{align}
and
\begin{align}
    p_1'\cdot q=p_2'\cdot q=p_1\cdot q=p_2\cdot q=\frac{q^2}{2},
\end{align}
we obtain
\begin{align}\label{eqn:sigmatotaleeeefree}
    \sigma_{\text{tot},1+1}^{e^+e^-\rightarrow \mu^+\mu^-}=\frac{2e^4m_e^2m_{\mu}^2}{q^8}\frac{1}{\sqrt{1-\frac{4m_e^2}{q^2}}\sqrt{1-\frac{4m_{\mu}^2}{q^2}}}\Theta(q^2-4m_{\mu}^2).
\end{align}
However, this cross-section in 1+1D is apparently divergent when $q^2$ approaches the threshold of the $\mu^+ \mu^-$  pair production, $q^2 = 4m_{\mu}^2$. Due to the two missing dimensions in contrast to the 3+1 dimensions, the factor $\sqrt{1-\frac{4m_{\mu}^2}{q^2}}$ here is positioned on the denominator rather than the numerator.
This corresponds to the singular behavior of the vacuum polarization function $T(q^2)$ at $q^2=4m^2$ as we noticed earlier. We have interpreted the singular behavior at $q^2=4m^2$ as the symptom of the confinement in QED(1+1), namely a fermion or anti-fermion cannot be on its mass shell, \'a la ``Schwinger mechanism''.
However, the singularity of the vacuum polarization amplitude $T(q^2)$ at $q^2=4m^2$ is removed
in the geometric sum to yield the 
dressed photon propagator given by
Eq.~(\ref{eqn:photon_dressedprop}).
Replacing the bare photon propagator by the dressed photon propagator, 
we get
\begin{align}
&\sigma_{\text{tot,1+1}}^{e^+e^-\rightarrow \mu^+\mu^-}\notag\\
=&\frac{2e^4m_e^2m_{\mu}^2}{\left|q^2+q^2T_{\mu^+\mu^-}(q^2)\right|^2}\frac{1}{q^4}\frac{1}{\sqrt{1-\frac{4m_e^2}{q^2}}}\frac{1}{\sqrt{1-\frac{4m_{\mu}^2}{q^2}}}\Theta(q^2-4m_{\mu}^2).
\end{align}
Now, plugging in the self-energy of the photon dressed by muon loops, $T_{\mu^+\mu^-}(q^2)$, i.e. 
$T(q^2)$ with $m=m_\mu$, we have
\begin{align}\label{eqn:sigmatotaleeeefull}
    &\sigma_{\text{tot,1+1}}^{e^+e^-\rightarrow \mu^+\mu^-}\notag\\
    =&\frac{2e^4m_e^2m_{\mu}^2}{\left| q^2-\frac{e^2}{\pi}\left[1-\frac{m_{\mu}^2/q^2}{\sqrt{1/4-m_{\mu}^2/q^2}}\ln\left(\frac{1-\frac{1}{2\sqrt{1/4-m_{\mu}^2/q^2}}}{1+\frac{1}{2\sqrt{1/4-m_{\mu}^2/q^2}}}\right)\right]\right|^2}\frac{1}{q^4}\frac{1}{\sqrt{1-\frac{4m_e^2}{q^2}}}\notag\\
    &\cdot\frac{1}{\sqrt{1-\frac{4m_{\mu}^2}{q^2}}}\Theta(q^2-4m_{\mu}^2)\notag\\
    =&2e^4m_e^2m_{\mu}^2\cdot \left[ \left(q^2-\frac{e^2}{\pi}\right)\sqrt{1-\frac{4m_e^2}{q^2}}\right.\notag\\
    &\left.+\frac{2e^2m_\mu^2}{\pi q^2}\frac{\sqrt{q^2-4m_e^2}}{\sqrt{q^2-4m_{\mu}^2}}\ln\left(\frac{1-\frac{1}{2\sqrt{1/4-m_{\mu}^2/q^2}}}{1+\frac{1}{2\sqrt{1/4-m_{\mu}^2/q^2}}}\right)\right]^{-1}\notag\\
    &\cdot \left[ \left(q^2-\frac{e^2}{\pi}\right)\sqrt{1-\frac{4m_{\mu}^2}{q^2}}+\frac{2e^2m_{\mu}^2}{\pi q^2}\ln\left(\frac{1-\frac{1}{2\sqrt{1/4-m_{\mu}^2/q^2}}}{1+\frac{1}{2\sqrt{1/4-m_{\mu}^2/q^2}}}\right)\right]^{-1}\notag\\
    &\cdot\frac{1}{q^{4}}\ \Theta(q^2-4m_{\mu}^2),
\end{align}
which is no longer divergent at $q^2=4m_{\mu}^2$ due to the loop corrections. This phenomenon again indicates confinement, i.e., in a confining theory, a single particle propagating on its own, i,e, a bare propagator, is not well-defined, and only when one sums over the loop diagrams and uses the dressed propagator (like the constituent quark idea), can one obtain the physical quantities such as the total cross section not to diverge.
Taking $e=1$ and $m_e=m_{\mu}=1$ as an example, we plot $\sigma_{\text{tot},1+1}^{e^+e^-\rightarrow \mu^+\mu^-}(q^2)$ in Fig. \ref{fig:sigmatotaleeee}, where we can see the different behavior of Eqs. (\ref{eqn:sigmatotaleeeefree}) and (\ref{eqn:sigmatotaleeeefull}) near $q^2=4$ in that one diverges and the other does not.

Having computed the total cross section of the fermion and anti-fermion pair production from the $e^+ e^-$ annihilation process, we may now discuss the ratio of the total cross sections between the quarks and the leptons known as the $R$ ratio in 3+1D given by 
\begin{equation}\label{eqn:defRratio}
R_{3+1}=\frac{\sigma_{\text{tot},3+1}^{e^+e^-\rightarrow q\bar{q}}}{\sigma_{\text{tot},3+1}^{e^+e^-\rightarrow \mu^+\mu^-}}.
\end{equation}
Typically in the $R$ ratio, the quark masses are not neglected while the lepton masses are neglected for the discussion of the different quark flavor thresholds which have been measured from the experiments. 

For the lepton pair production cross section, neglecting both the electron and muon masses, 
we get the spin-averaged invariant amplitude square given by~\cite{Halzen}
\begin{equation}\label{eqn:M3+1}  |\overline{\mathcal{M}}_{3+1}|^2\longrightarrow 2e^4\frac{t^2+u^2}{s^2},
\end{equation}
where $s,t$ and $u$ are the usual Mandelstam variables. 
Noting that in the ultra-relativistic limit~\cite{Halzen}
\begin{align}
    -\frac{u}{s}&\longrightarrow\ \frac{1}{2}\left(1+\cos\theta\right)\label{eqn:3+1uovers}\\
    -\frac{t}{s}&\longrightarrow\ \frac{1}{2}\left(1-\cos\theta\right)\label{eqn:3+1tovers},
\end{align}
one can get 
\begin{align}
    &\sigma_{\text{tot},3+1}^{e^+e^-\rightarrow \mu^+\mu^-}
    =\frac{\alpha^2}{4\pi^2s}\int d\Omega\ \left(1+\cos^2\theta\right),
\end{align}
where $\alpha=\frac{e^2}{4\pi}$. 
Integrating over the solid angle gives the familiar result
\begin{align}\label{eqn:crosssectionleptons}
    \sigma_{\text{tot},3+1}^{e^+e^-\rightarrow \mu^+\mu^-}=\frac{4\pi\alpha^2}{3s}.
\end{align}

Similarly, the numerator of Eq.~(\ref{eqn:defRratio}) can be obtained by replacing the muon mass in the derivation of Eqs. (\ref{cross-section(3+1)}) through (\ref{cross-section(3+1)res}) with the quark masses $m_{q_i}$ and noting that the quark charges $Q_{q_i}$ are different from the electron and muon charge $e$. 
Using 
\begin{equation}
    p_1\cdot p_1'=p_2\cdot p_2'=\frac{m_{q_i}^2-t}{2}
\end{equation}
\begin{equation}
    p_1\cdot p_2'=p_2\cdot p_1'=\frac{m_{q_i}^2-u}{2}
\end{equation}
as well as Eq. (\ref{eqn:p1dotp2}) and Eqs. (\ref{eqn:3+1uovers})-(\ref{eqn:3+1tovers}),
we obtain 
\begin{align}\label{eqn:crosssectionqqbar}
    &\sigma_{\text{tot},3+1}^{e^+e^-\rightarrow q\bar{q}}\notag\\
    =&\sum_{q_i}\frac{1}{4\pi s}\sqrt{1-\frac{4m_{q_i}^2}{s}}Q_{q_i}^2e^2\left(\frac{1}{3}+\frac{m_{q_i}^4}{s^2}+2\frac{m_{q_i}^2}{s}\right)\Theta(q^2-4m_{q_i}^2).
\end{align}
The $R$ ratio in Eq. (\ref{eqn:defRratio}) is then given by 
\begin{align}\label{eqn:Rratio3+1result}
    R_{3+1}=3\sum_{q_i}\sqrt{1-\frac{4m_{q_i}^2}{q^2}}\frac{Q_{q_i}^2}{e^2}\left(1+3\frac{m_{q_i}^4}{s^2}+6\frac{m_{q_i}^2}{s}\right)\Theta(q^2-4m_{q_i}^2),
\end{align}
where the color factor 3 is taken into account.  

The analogous $R$ ratio in $1+1$ dimensions can then be computed from the cross section of $e^+e^-\rightarrow q_i\bar{q_i}$ for each quark flavor $q_i$ to that of $e^+e^-\rightarrow \mu^+\mu^-$ and sum them over all the quark flavors. In contrast to the cross sections in 3+1D, 
we need to take into account the overall mass factors appearing in Eq.~(\ref{eqn:sigmatotaleeeefull}). 
In the limit of both $m_e \to 0$ and $m_\mu \to 0$, 
the total cross section $\sigma_{\text{tot}}^{e^+e^-\rightarrow \mu^+\mu^-}$ scaled by $m_e$ and $m_\mu$ in 1+1D can be obtained from Eq. (\ref{eqn:sigmatotaleeeefull}) as
\begin{align}
     \frac{\sigma_{\text{tot},1+1}^{e^+e^-\rightarrow \mu^+\mu^-}}{m_e^2m_{\mu}^2}= \frac{2e^4 }{q^4 \left( q^2-\frac{e^2}{\pi }\right) ^2}\Theta (q^2).\label{eqn:sigmatotignoremasses}
\end{align}
Similarly, we may scale the individual total cross section
for each quark flavor $q_i$ denoted by $\sigma_{\text{tot},1+1}^{e^+e^-\rightarrow q_i\bar{q_i}}$ by $m_e$ and $m_{q_i}$ replacing $m_\mu$ by $m_{q_i}$ in Eq. (\ref{eqn:sigmatotaleeeefull}) and obtain
\begin{align}
\label{1+1Dcross-section-for-quark-flavor}
       \frac{\sigma_{\text{tot},1+1}^{e^+e^-\rightarrow q_i\bar{q_i}}}{m_e^2 m_{q_i}^2} =&\frac{2Q_{q_i}^2e^2}{\left| q^2-\frac{Q_{q_i}^2}{\pi}\left[1-\frac{m_{q_i}^2/q^2}{\sqrt{1/4-m_{q_i}^2/q^2}}\ln\left(\frac{1-\frac{1}{2\sqrt{1/4-m_{q_i}^2/q^2}}}{1+\frac{1}{2\sqrt{1/4-m_{q_i}^2/q^2}}}\right)\right]\right|^2}\notag\\
       &\cdot\frac{1}{q^4}\frac{1}{\sqrt{1-\frac{4m_e^2}{q^2}}}\frac{1}{\sqrt{1-\frac{4m_{q_i}^2}{q^2}}}\Theta(q^2-4m_{q_i}^2),
\end{align}
where we did not neglect any mass although we will take practically $m_e = 0$ in our numerical computations. Summing over the quark flavors in Eq.~(\ref{1+1Dcross-section-for-quark-flavor}) and taking its ratio to Eq.~(\ref{eqn:sigmatotignoremasses}), we may define the R ratio in 1+1D as  
\begin{equation} 
R_{1+1}=\frac{3\sum_{q_i} \frac{\sigma_{\text{tot},1+1}^{e^+e^-\rightarrow q_i\bar{q_i}}}{m_e^2m_{q_i}^2}}{\frac{\sigma_{\text{tot},1+1}^{e^+e^-\rightarrow \mu^+\mu^-}}{m_e^2m_{\mu}^2}}, 
\end{equation}
where the color factor 3 is taken into account.

\begin{figure}
	\centering
\includegraphics[width=0.95\linewidth]{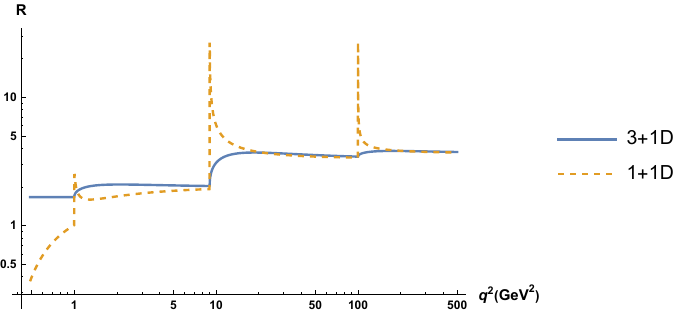}
	\caption{The $R$ ratio in 1+1D and 3+1D plotted as a function of $q^2$.}
	\label{fig:Rratioplot3}
\end{figure}

To contrast the behaviors of $R_{1+1}$ and $R_{3+1}$ near the thresholds of
different quark flavors, we plot them in Fig.~\ref{fig:Rratioplot3} taking $e=1$, the masses of the heavier quarks as $m_s= 0.5\ \text{GeV}$, $m_c= 1.5\ \text{GeV}$, and $m_b= 5\ \text{GeV}$ and ignoring the $u,d$ quark as well as the electron and muon masses.
While we can clearly see the thresholds at $1,\ 9$, and $100\ \text{GeV}^2$, which are the $4m_s^2$, $4m_c^2$, and $4m_b^2$, respectively, we note that the two $R$ ratios have quite different behavior near these thresholds. In 3+1D, a non-confining theory, the step functions rise smoothly, while in 1+1D, a confining theory, we observe these spikes at the thresholds, which would shoot to infinity had we not used the dressed propagator summing over all one-loop diagrams instead of just the lowest order result.\\

\section{\label{sec:conclusion}Conclusions and outlook}

In this work, we studied the QED vacuum polarizations  of 1+1D vs. 3+1D to investigate the nature of threshold singularity, ultraviolet divergence and regularization depending on the spacetime where the corresponding nontrivial Dirac vacuum exists. We found that our results both in 1+1D and 3+1D satisfy the dispersion relation (DR), respectively, assuring their validity supported by the first principle  quantum field theory. 

We note that the 1+1D QED vacuum polarization exhibits the fermion threshold singularity at the one-loop level, which may be interpreted as the symptom of the fermion confinement in 1+1D, namely no isolated fermion can be found in 1+1D, {\it 'a la} Schwinger mechanism. While the geometric sum of the one-loop vacuum polarization amplitudes shown in Fig.~\ref{fig:Photon_dress} regulates the divergence appearing in the single loop amplitude, the regularized result comes with the mass generation of the gauge field. When the photon gets massive in 1+1D QED, it acquires the longitudinal polarization as there are no transverse polarizations of the photon in 1+1D. 

This indicates the correlation between the charge confinement and the photon mass generation in 1+1D. 
As illustrated in Eqs.~(\ref{axial-anomaly})-(\ref{axial-anomaly-result}), the photon mass generation is also accompanied by the axial anomaly in 1+1D. Thus, in 1+1D, the charge confinement and the chiral symmetry breaking appear tied together with the photon mass generation. 
This is quite remarkable from the QCD perspectives,
as the color confinement and chiral symmetry breaking are still regarded as two distinguished and separate phenomena. While the QCD vacuum structure studies are ongoing vigorously to describe both phenomena together~\cite{Shuryak:2021vnj}, it seems to call for more in-depth study on the underpinning mechanism to understand both the charge confinement and the chiral symmetry breaking together within the common relativistic quantum field theoretic framework. Further studies on the time-ordered processes of the fermion and anti-fermion pair creation and annihilation are underway contrasting the two distinguished forms of relativistic dynamics proposed by Dirac~\cite{Dirac:1949cp}, namely the instant form dynamics and the light-front dynamics.

\section*{Acknowledgement}

This work was supported by the U.S. Department of Energy (Grant No. DE-FG02-03ER41260). 
The National Energy Research Scientific Computing Center (NERSC) supported by the Office of Science of the U.S. Department of Energy 
under Contract No. DE-AC02-05CH11231 is also acknowledged. Bailing Ma was supported by the U.S. Department of Energy, Office of Science, Office of Nuclear Physics, under contract no. DE-AC02-06CH11357 during her employment at Argonne National Laboratory.

\appendix
\section{\label{sec:app}Ultra-relativistic Amplitude Squares in 1+1D and 3+1D}

We contrast the difference of the
total cross sections given by Eqs.~(\ref{eqn:sigmatotaleeeefree}) and (\ref{eqn:crosssectionleptons}),  respectively, in 1+1D and 3+1D.  
In the massless limit or the ultra-relativistic limit, we note that the total cross section in 1+1D given by Eq.~(\ref{eqn:sigmatotaleeeefree}) vanishes, while in the same limit the total cross section in 3+1D given by 
Eq.~(\ref{eqn:crosssectionleptons}) does not vanish. This difference can be understood from the difference in the amplitude squares in 1+1D and 3+D. Following the derivations of the total cross section in 1+1D illustrated in Eqs.~(\ref{sigma}) - (\ref{eqn:sigmatotaleeeefree}), we note that the amplitude square in 1+1D in the ultra-relativistic limit is given by 
\begin{equation}
|\overline{\mathcal{M}}_{1+1}|^2\longrightarrow \frac{e^4}{2}\frac{t^2+u^2-s^2}{s^2}=\frac{e^4}{2}\frac{-2tu}{s^2},
\end{equation}
while the corresponding amplitude square in 3+1D is given by Eq.~(\ref{eqn:M3+1}).
We note that either Eq. (\ref{eqn:3+1uovers}) or Eq. (\ref{eqn:3+1tovers}) is zero because $\theta$ is either $0$ or $\pi$ and thus $|\overline{\mathcal{M}}_{1+1}|^2$ vanishes. 
This is in agreement with the result we obtained in Eq. (\ref{eqn:sigmatotaleeeefree}), when $m_e,m_{\mu}\longrightarrow 0$, $\sigma_{\text{tot},1+1}^{e^+e^-\to\mu^+\mu^-}\longrightarrow 0$.
We attribute this difference to the absence of transverse polarization of the virtual photon in 1+1D, as we show explicitly below in this Appendix.   

If we calculate different polarization of the photon contribution to $|\overline{\mathcal{M}}_{3+1}|^2$, using
\begin{align}
    p_1'&=(m_{\mu},0,0,0)\notag\\
    p_2'&=(\frac{q^2-2m_{\mu}^2}{2m_{\mu}},0,0,\frac{\sqrt{q^2}\sqrt{q^2-4m_{\mu}^2}}{2m_{\mu}})\notag\\
    q&=(\frac{q^2}{2m_{\mu}},0,0,\frac{\sqrt{q^2}\sqrt{q^2-4m_{\mu}^2}}{2m_{\mu}})
\end{align}
and
\begin{align}
    \varepsilon_{(0)}&=(\frac{\sqrt{q^2-4m_{\mu}^2}}{2m_{\mu}},0,0,\frac{\sqrt{q^2}}{2m_{\mu}})\notag\\
    \varepsilon_{(+1)}&=-\frac{1}{\sqrt{2}}(0,1-i,1+i,0)\notag\\
    \varepsilon_{(-1)}&=\frac{1}{\sqrt{2}}(0,1+i,1-i,0),
\end{align}
the contribution of longitudinally polarized photon to the amplitude squared is
\begin{align}
    &|\overline{\mathcal{M}}_{3+1,\lambda=0}|^2\notag\\
    =&\frac{4e^4}{q^4}\left[\frac{s}{2}+m_{\mu}^2-2\left(\varepsilon_{(0)}\cdot p_1'\right)^2\right]\left[\frac{s}{2}+m_{e}^2-2\left(\varepsilon_{(0)}\cdot p_1\right)^2\right]\notag\\
    =&\frac{4e^4}{q^4}\left[\frac{q^2}{2}+m_{\mu}^2-\frac{q^2-4m_{\mu}^2}{2}\right]\left[\frac{q^2}{2}+m_{e}^2-\frac{q^2-4m_{e}^2}{2}\right].
\end{align}
In the limit of $m_e,m_{\mu}\longrightarrow 0$, this is 0. On the other hand,
\begin{align}
    &|\overline{\mathcal{M}}_{3+1,\lambda=\pm 1}|^2\notag\\
    =&\frac{4e^4}{q^4}2\left[\frac{s}{2}+m_{\mu}^2+2\left(\varepsilon_{(+1)}\cdot p_1'\right)\left(\varepsilon_{(-1)}\cdot p_1'\right)\right]\notag\\
    &\cdot \left[\frac{s}{2}+m_{e}^2+2\left(\varepsilon_{(+1)}\cdot p_1\right)\left(\varepsilon_{(-1)}\cdot p_1\right)\right]\notag\\
    =&\frac{4e^4}{q^4}2\left[\frac{q^2}{2}+m_{\mu}^2\right]\left[\frac{q^2}{2}+m_{e}^2\right].
\end{align}
In the limit of $m_e,m_{\mu}\longrightarrow 0$, this becomes $2e^4$, which is exactly equal to Eq. (\ref{eqn:M3+1}) in the forward/backward direction. Thus, the forward/backward annihilation cross section in the ultra-relativistic limit in 3+1D is completely coming from the transverse polarization of the photon, which explains the reason why it vanishes in 1+1D.

\bibliographystyle{elsarticle-num} 
\bibliography{example}






\end{document}